\documentclass[12pt,fleqn]{article}
\usepackage{epsfig}
\usepackage[dvips]{color}
\begin{document}
\title{Spontaneous CP violation in the triplet extended supersymmetric standard model}
\author{S. W. Ham$^{(1)}$\footnote{swham@knu.ac.kr} and S. K. Oh$^{(2,3)}$\footnote{sunkun@konkuk.ac.kr}
\\
\\
{\it $^{(1)}$ Department of Physics, KAIST, Daejon 305-701, Korea}
\\
{\it $^{(2)}$ Department of Physics, Konkuk University, Seoul 143-701, Korea}
\\
{\it $^{(3)}$ Center for High Energy Physics, Kyungpook National University}
\\
{\it Daegu 702-701, Korea}
\\
\\
}
\date{}
\maketitle
\begin{abstract}
We find that, at the one-loop level, the spontaneous CP violation is possible in a
supersymmetric standard model that has an extra chiral Higgs triplet with hypercharge $Y=0$.
At the tree level, this triplet-extended supersymmetric standard model (TESSM) cannot
have any reasonable parameter spaces for the spontaneous CP violation,
because the experimental constraints on the coupling coefficient of the neutral Higgs boson
to a pair of $Z$ bosons exclude them.
By contrast, at the one-loop level, we find that there are experimentally allowed parameter regions,
where the spontaneous CP violation may take place.
The mass of the lightest neutral Higgs boson in the TESSM in this case may be as large as about 100 GeV,
by considering the one-loop contribution due to the top quark and squark loops.
\end{abstract}

\vfil\eject

\section{Introduction}

It is well known that there are some essential issues for which the Standard Model (SM) is insufficient
to afford satisfying answers.
One of them is the naturalness problem, which states that the mass of the Higgs boson in the SM requires
a fine tuning in order to remain in the range of the electroweak scale,
because it receives the quadratic divergence from radiative corrections due to the SM particle loops.
The supersymmetry can solve the naturalness problem as it stabilizes the Higgs boson mass by virtue of the superpartners.
Not only the naturalness problem, but also the gauge coupling unification, the gauge hierarchy problem,
as well as the possibility of incorporating the gravity by the local supersymmetry, can be addressed
if the nature is supersymmetric.

Since the advent of the supersymmetry several decades ago, the formalism and
the characteristic properties have well been established and accepted
into the theoretical mainstream of high energy physics.
Nowadays, the supersymmetry is widely regarded as the one of the key ingredients
that the new physics beyond the SM should possess.
The phenomenology of supersymmetric models has extensively been studied in recent years,
in anticipation that the Large Hadron Collider would certainly provide the first clues
for the existence of the supersymmetry.
For the comprehensive reviews on the pioneering studies on the supersymmetry, we refer to Ref. [1].

The simplest version of the supersymmetrically extended SM is called the minimal supersymmetic standard model (MSSM),
which has just two Higgs doublets in order to give masses independently to the up-type and down-type fermions.
In the MSSM, the mass of its lightest scalar Higgs boson is predicted to be smaller than the $Z$ boson mass
at the tree level, since the quartic coupling is given only by the weak gauge couplings.
This small tree-level mass is rejected phenomenologically by the negative experimental results.
At the one-loop level, while the lightest scalar Higgs boson in the MSSM may receive
large radiative corrections due to the top and stop quark loops to become heavy enough
to satisfy the experimental lower bound, the same radiative corrections increase
too much the quadratic term of the Higgs potential.
Hence, the little hierarchy problem in the MSSM.
In addition, the MSSM has the well-noticed $\mu$-parameter problem [2].

In order to alleviate the shortcomings of the MSSM and/or to explore further theoretical frontiers,
various nonminimal versions of the supersymmetrically extended SM have been introduced in the literature.
The most well-known among them is the next-to-minimal supersymmetric standard model (NMSSM),
which possesses an extra Higgs singlet besides the two Higgs doublets of the MSSM.
The NMSSM provides a plausible solution for the $\mu$-parameter problem and reduce
the burden of fine tuning for the little hierarchy problem of the MSSM.
Moreover, in the NMSSM, the lightest scalar Higgs boson may be heavier than the Z boson even
at the tree level because there is an additional quartic coupling [3].

As another nonminimal version of the supersymmetrically extended SM,
we have the triplet extended supersymmetric model (TESSM).
It has an additional complex Higgs triplet besides the two Higgs doublets of the MSSM,
and has been proposed some years ago in order to bring about an explicit breaking of the custodial SU(2) symmetry
and push up the mass of the lightest scalar Higgs boson in the MSSM [4].
The Higgs triplets may be found not only in the TESSM but also in other models, for example,
in a model for massive neutrinos, embedded in a unified gauge group [5].

Recently, the interest in the TESSM has been revived.
In the literature, many studies on the Higgs sector of the TESSM can be found,
including the calculation of the upper bound on the mass of the lightest neutral Higgs boson,
the examination of the probability of producing them in $e^+e^-$ collisions at the ILC,
the phenomenology of charged Higgs bosons as well as neutral Higgs bosons,
and calculation of one-loop level radiative corrections in the Higgs sector [6,7,8,9].
These studies show that the Higgs sector of the TESSM may provide new possibilities,
as an alternative to the NMSSM.

Concerning the CP violation, the SM is found to be inadequate to produce
the right amount of CP violation for baryogenesis [10].
In the MSSM, investigations have revealed that neither spontaneous nor explicit CP violation may
be realized at the tree level, and that the explicit CP violation but not the spontaneous
one is possible in the radiatively corrected Higgs sector of the MSSM [11].
In the NMSSM, there is some difficulty to produce the spontaneous CP violation at the tree level [12],
but the tree-level explicit CP violation is allowed [13].
We have studied elsewhere the possibility of the explicit CP violation in the NMSSM at the one-loop level [14].

In this article, we would like to study the TESSM with respect to the CP violation.
In particular, we are interested in the possibility of spontaneous CP violation in the TESSM.
In order to bring about the spontaneous CP violation, we assume that the Higgs potential of the TESSM
may develop complex vacuum expectation values when the electroweak symmetry is broken.
We study the effects of these complex phases on the masses of the neutral Higgs bosons in the TESSM
and on the coupling coefficients of the neutral Higgs bosons to a pair of $Z$ bosons,
at the tree level as well as at the one-loop level.
As the recent report on the LEP experiments set a model-independent constraints
on the Higgs-$Z$-$Z$ coupling coefficients, one can examine whether or not the scenario
of spontaneous CP violation in the TESSM may be allowed by LEP experiments [15].

In this article, we report the result of our study that, for a reasonable parameter space,
when radiative corrections due to top and stop quark loops are taken into account,
the spontaneous CP violation is quite possible in the TESSM at the one-loop level.
However, we find that the LEP experiments rule out the possibility of tree-level spontaneous CP violation
in the TESSM.

\section{Higgs sector}

The most general renormalizable, gauge-invariant superpotential for the interactions
among the Higgs superfields in the TESSM is given as [4]
\begin{equation}
{\cal W} = \lambda {\cal H}_1 \epsilon \Sigma {\cal H}_2 + \mu_D {\cal H}_1 \epsilon {\cal H}_2
+ \mu_T {\rm Tr} (\Sigma^2)
\end{equation}
where ${\cal H}_1$ and ${\cal H}_2$ are doublet Higgs superfields, $\Sigma$ is
chiral complex triplet Higgs superfield, $\epsilon$ is an antisymmetric $2 \times 2$ matrix with $\epsilon_{12} = 1$,
and three parameters $\lambda$, $\mu_D$, and $\mu_T$.
While $\lambda$ is dimensionless, both $\mu_D$ and $\mu_T$ are of mass dimension.
Thus, the so-called $\mu$-parameter problem in the MSSM is not solved in the TESSM.
In the TESSM, these $\mu$ parameters trigger the right size of the electroweak symmetry breaking.

The Higgs sector of the TESSM consists of two Higgs doublets $H_1$, $H_2$ and
a Higgs triplet $\Sigma$ (denoted by the same notation as the corresponding superfield,
but distinguishing between them would be straightforward), which may be expressed as
\begin{equation}
H_1 = \left(
\begin{array}{c}
H_1^0 \\
H_1^-
\end{array}
\right) \ , \quad
H_2= \left(
\begin{array}{c}
H_2^+ \\
H_2^0
\end{array}
\right)  \ , \quad
\Sigma = \left(
\begin{array}{cc}
{\displaystyle  {1 \over \sqrt{2}} } \xi^0 & - \xi_2^+ \\
\xi_1^- & {\displaystyle  - {1 \over \sqrt{2}} } \xi^0
\end{array}
\right) \ ,
\end{equation}
where $\xi^0$ is the neutral Higgs field and $\xi_1^-$, $\xi_2^+$ are
the charged Higgs fields of the complex Higgs triplet.
Note that $\xi_1^- \neq  -(\xi_2^+)^*$ unlike the case of a real Higgs triplet
in a non-supersymmetric version [4].
By contrast, $H_1^- = (H_2^+)^*$.
The hypercharges for $H_1$, $H_2$ are respectively $Y=-1/2$ and $Y=1/2$, and
the hypercharge for $\Sigma$ is $Y=0$.
The physical Higgs boson family in the TESSM after electroweak symmetry breaking consists of
three charged Higgs bosons and five neutral Higgs bosons.

The Higgs potential in the TESSM may be constructed by collecting relevant terms from the superpotential.
The resulting Higgs potential, $V_0$, at the tree level, may be written as
\begin{equation}
V_0 = V_F + V_D + V_{\rm S} \ ,
\end{equation}
where
\begin{eqnarray}
V_F & = &  \bigg| \mu_D H_2^0 + \lambda \bigg( H_2^+ \xi^-_1 - {1 \over \sqrt{2}} H_2^0 \xi^0 \bigg) \bigg|^2
+ \bigg| \mu_D H_1^0 + \lambda \bigg( H_1^- \xi^+_2 - {1 \over \sqrt{2}} H_1^0 \xi^0 \bigg) \bigg|^2  \cr
& &\mbox{} + \bigg| \mu_D H_2^+ + \lambda \bigg( {1 \over \sqrt{2}} H_2^+ \xi^0 - H_2^0 \xi_2^+ \bigg) \bigg|^2
+ \bigg| \mu_D H_1^- + \lambda \bigg( {1 \over \sqrt{2}} H_1^- \xi^0 - H_1^0 \xi^-_1 \bigg) \bigg|^2   \cr
& &\mbox{} + \bigg|2 \mu_T \xi^0 - {\lambda \over \sqrt{2}} \bigg( H_1^0 H^0_2 + H_1^- H^+_2 \bigg) \bigg|^2
+ \bigg| \lambda  H_1^0 H^+_2 - 2 \mu_T \xi_2^+ \bigg|^2        \cr
& &\mbox{} + \bigg| \lambda H_1^- H^0_2 - 2 \mu_T \xi_1^- \bigg|^2 \ , \cr
V_{D} & = & {g_2^2 \over 8} \bigg[ |H_1^0|^2 - |H_1^-|^2 + |H^+_2|^2 - |H_2^0|^2 + 2 |\xi^+_2|^2 - 2|\xi_1^-|^2 \bigg]^2  \cr
& &\mbox{} + {g_1^2 \over 8} \bigg[ |H_1^0|^2 + |H_1^-|^2 - |H^+_2|^2 - |H_2^0|^2  \bigg]^2    \cr
& &\mbox{} + {g_2^2 \over 8} \bigg[ H_1^{0*} H_1^- + H^{+*}_2 H_2^0 + \sqrt{2} ( \xi^+_2 + \xi_1^- ) \xi^{0*} + {\rm H.c.} \bigg]^2 \cr
& &\mbox{} - {g_2^2 \over 8} \bigg[ H_1^{-*} H_1^0 + H^{0*}_2 H_2^+ + \sqrt{2} ( \xi^+_2 - \xi_1^- ) \xi^{0*} - {\rm H.c.} \bigg]^2 \ , \cr
V_{\rm S} & = & m_1^2 |H_1|^2 + m_2^2 |H_2|^2 + m_3^2 {\rm Tr} (\Sigma^\dag \Sigma)   \cr
& &\mbox{} + [A_{\lambda} \lambda H_1 \epsilon \Sigma H_2 + B_D \mu_D H_1 \epsilon H_2 + B_T \mu_T {\rm Tr} (\Sigma^2) + {\rm H.c.} ] \ .
\end{eqnarray}
with $g_1$ and $g_2$ being, respectively, the $U(1)$ and $SU(2)$ gauge coupling coefficients,
$A_{\lambda}$ the trilinear parameter, $B_D$ and $B_T$ the bilinear parameters,
and $m_i$ ($i=1,2,3$) the soft SUSY breaking masses.
These soft SUSY breaking masses may later be eliminated by means of the three minimization conditions
with respect to the three neutral Higgs fields of the Higgs doublets and the Higgs triplet.

The neutral Higgs potential, $V_N$, responsible for the electroweak symmetry breaking, may be obtained from $V_0$.
Explicitly, it is written as
\begin{eqnarray}
V_{\rm N} & = & \bigg| \mu_D H_2^0 - {\lambda \over \sqrt{2}} H_2^0 \xi^0 \bigg|^2
+ \bigg| \mu_D H_1^0 - {\lambda \over \sqrt{2}}  H_1^0 \xi^0 \bigg|^2  \cr
& &\mbox{} + \bigg|2 \mu_T \xi^0 - {\lambda \over \sqrt{2}} H_1^0 H^0_2 \bigg|^2
+ { g_1^2 + g_2^2 \over 8} \bigg [ |H_1^0|^2 - |H_2^0|^2  \bigg]^2 \cr
& &\mbox{} + \bigg[ - {\lambda \over \sqrt{2}} A_{\lambda}  H_1^0 H_2^0 \xi^0
+ B_D \mu_D H_1^0 H_2^0 + B_T \mu_T \xi^0 \xi^0 + {\rm H.c.} \bigg] \  \cr
& &\mbox{} + m_1^2 |H_1^0|^2 + m_2^2 |H_2^0|^2 + m_3^2 |\xi^0|^2 \ .
\end{eqnarray}
We assume that all the free parameters in $V_N$ are real such that the CP symmetry would not be broken explicitly.
However, we assume that the CP symmetry may spontaneously broken by means of the complex vacuum expectation values
of the neutral Higgs fields.
Thus, without loss of generality, we may take the vacuum expectation values of the neutral Higgs fields as
$v_1 = \langle H_1^0 \rangle$, $v_2 e^{\phi_1} = \langle H_2^0 \rangle$,
and $x e^{\phi_2} = \langle \xi^0 \rangle$ ($v_1$, $v_2$, and $x$ are of course real).
Note that between $\langle H_1^0 \rangle$ and $\langle H_2^0 \rangle$
we can make one of them real because the physically meaningful phase is not
their individual phases but the relative phase between them.

The size of the vacuum expectation value of $\xi^0$ in the TESSM is known
to receive a strong experimental constraint from the $\rho$-parameter, which is expressed in the TESSM as
\begin{equation}
\rho \equiv {m_W^2 \over m_Z^2 \cos^2 \theta_W} = 1 + 4 {x^2 \over v^2}  \ ,
\end{equation}
where $v^2 = v_1^2 + v_2^2$ and
$m_W^2 = g_2^2 (v^2 + 4 x^2) /2$ and $m_Z^2 = g_2^2 v^2 /(2 \cos^2 \theta_W$) are
respectively the squared masses of charged and neutral weak gauge bosons.
Experimental results for the $\rho$ parameter impose an upper bound on $x/v \le 0.012$,
which in turn constrains $x < 3$ GeV [9].

In order to study the one-loop contributions, we employ the effective potential method [16], which gives
the one-loop effective potential as
\begin{equation}
V_1  = \sum_{k} {n_k {\cal M}_k^4 \over 64 \pi^2}
\left [
\log {{\cal M}_k^2 \over \Lambda^2} - {3 \over 2}
\right ]  \ ,
\end{equation}
where $\Lambda$ is the renormalization scale in the modified minimal subtraction scheme,
and $n_k$ are the degrees of freedom from color, charge, and spin factors of the particles
that enter into the loops.
We only take into account the top and stop quark loops since their contributions are most dominant
for most of the parameter space, though for very large values of $\tan \beta$ such as 50 the bottom
and sbottom quark loops may also give phenomenologically significant contributions in low energy supersymmetric models.
Thus, we have $n_t = -12$ for top quarks and $n_{{\tilde t}_i} = 6$ ($i=1,2$) for the stop quarks.

The top quark mass is given by $H_2$ as $m_t = h_t v_2$, where $h_t$ is the Yukawa coupling coefficient
for the top quark, and the masses of the stop quarks are obtained as
\begin{equation}
m_{{\tilde t}_1, {\tilde t}_2}^2 = {m_Q^2 + m_U^2 \over 2} + h_t^2 v_2^2
+ {g_1^2 + g_2^2 \over 8} (v_1^2 - v_2^2) \mp \sqrt{X_t} \ ,
\end{equation}
where $m_Q$ and $m_U$ are the soft SUSY breaking masses for the stop quarks,
$A_t$ is the trilinear soft SUSY breaking parameters for the stop quarks,
and $X_t$ represents the stop quark mixing.

It is from $X_t$ that the possibility of the spontaneous CP violation arises.
In terms of Higgs fields, $X_t$ is explicitly given as
\begin{eqnarray}
X_t & = & \bigg [ {m_Q^2 - m_U^2 \over 2}
+ \bigg ( {g_2^2 \over 4} - {5 g_1^2 \over 12} \bigg ) \bigg (|H_1^0|^2 - |H_2^0|^2 \bigg)
   \bigg ]^2  \cr
& &\mbox{}   +  h_t^2 \bigg |A_t H_2^{0 *} + \lambda H_1^0 \xi^0/\sqrt{2} - \mu_D H_1^0 \bigg |^2  \ ,
\end{eqnarray}
which may possess complex phases when $H^0_2$ and $\xi^0$ develop complex vacuum expectation values.

Now, we would like to study two scenarios: In one scenario, where all the vacuum expectation values are
real and thus no complex phases contaminate the stop quark masses, CP is conserved.
The other scenario is our main subject, the CP-violating scenario, where complex phases
in the vacuum expectation values eventually trigger the scalar-pseudoscalar Higgs mixings
and thus the spontaneous CP violation.

\subsection{CP-conserving scenario}

Let us consider the CP-conserving scenario first, where we assume that
$\phi_1 = \phi_2 = 0$ in the vacuum expectation values of $H^0_2$ and $\xi^0$.
In this scenario, $X_t$ in the masses of the stop quarks are given as
\begin{eqnarray}
X_t & = & \left [ {m_Q^2 - m_U^2 \over 2} + \left ( {2 m_W^2 \over 3} - {5 m_Z^2 \over 12} \right ) \cos 2 \beta
   \right ]^2
   +  m_t^2 \bigg [(\mu_D \cot \beta - A_t)^2               \cr
& &\mbox{}  + \lambda x \cot \beta (\sqrt{2} A_t - \sqrt{2} \mu_D \cot \beta + { 1 \over 2} \lambda x \cot \beta ) \bigg ]  \ .
\end{eqnarray}

The five neutral Higgs bosons in this scenario have definite CP parities and may be divided
into CP-even and CP-odd states.
Two pseudoscalar Higgs bosons, $P_1$ and $P_2$, are constructed from
three imaginary components of neutral Higgs fields ${\rm Im} H^0_1$, ${\rm Im} H^0_2$,
and ${\rm Im} \xi^0$, among which a linear combination of ${\rm Im} H^0_1$
and ${\rm Im} H^0_2$ is gauged away.
On the basis of ($\sin\beta {\rm Im}H^0_1 + \cos\beta {\rm Im} H^0_2$, ${\rm Im} \xi^0$),
where $\tan\beta = v_2 /v_1$, $M_P$ is given as
\begin{equation}
M_P = \left ( \begin{array}{cc}
    M_{P11} & M_{P12}    \cr
    M_{P12} & M_{P22}
        \end{array}
    \right ) \ ,
\end{equation}
where, at the one-loop level,
\begin{eqnarray}
M_{P11} & = & - {2 B_D \mu_D \over \sin 2 \beta} + {\sqrt{2} \lambda A_{\lambda} x \over \sin 2 \beta}
+ {2 \sqrt{2} \lambda x \mu_T \over \sin 2 \beta}
- {3 m_t^2 A_t (2 \mu_D - \sqrt{2} \lambda x) \over 32 \pi^2 v^2 \sin^3 \beta \cos \beta}
f(m_{{\tilde t}_1}^2, m_{{\tilde t}_2}^2)     \   , \cr
M_{P22} & = &\mbox{} - 4 B_T \mu_T + {\lambda v^2 \mu_D \over \sqrt{2} x}
+ {\lambda A_{\lambda} v^2 \sin 2 \beta \over 2 \sqrt{2} x}
+ {\lambda v^2 \mu_T \sin 2 \beta \over \sqrt{2} x} \cr
& &\mbox{} - {3 \sqrt{2} m_t^2 \lambda (\mu_D \cot \beta - A_t) \over 32 \pi^2 x \tan \beta}
f(m_{{\tilde t}_1}^2, m_{{\tilde t}_2}^2)     \   , \cr
M_{P12} & = & {\lambda A_{\lambda} v \over \sqrt{2}}
- \sqrt{2} \lambda v \mu_T
+ {3 \sqrt{2} m_t^2 \lambda A_t \over 32 \pi^2 v \sin^2 \beta}
f(m_{{\tilde t}_1}^2, m_{{\tilde t}_2}^2)   \ ,
\end{eqnarray}
with
\begin{equation}
 f(m_x^2, \ m_y^2) = {1 \over (m_y^2 - m_x^2)} \left[  m_x^2 \log {m_x^2 \over \Lambda^2} - m_y^2
\log {m_y^2 \over \Lambda^2} \right] + 1 \ ,
\end{equation}
representing the radiative corrections.

The squared masses of these pseudoscalar Higgs bosons, $m^2_{P_1}$ and $m^2_{P_2}$,
are obtained from the symmetric $2\times 2$ mass matrix $M_P$ for the pseudoscalar Higgs bosons as
\begin{equation}
m_{P_1,P_2}^2 = {1 \over 2} \bigg[{\rm Tr} (M_P) \mp \sqrt{({\rm Tr} M_P)^2 - 4 {\rm det} (M_P)} \bigg] \ .
\end{equation}
These pseudoscalar Higgs bosons are sorted such that $m_{P_1} < m_{P_2}$.

The three scalar Higgs bosons, $S_i$ ($i = 1, 2, 3$), are constructed from
three real components of neutral Higgs fields ${\rm Re} H^0_1$, ${\rm Re} H^0_2$, and ${\rm Re} \xi^0$.
Their squared massed, $m_{S_i}$ ($i = 1, 2, 3$), are given as the eigenvalues of
the symmetric $3\times 3$ mass matrix $M_S$ for the scalar Higgs bosons.
These scalar Higgs bosons are sorted such that $m_{S_1} < m_{S_2} < m_{S_3}$.

Expressing $M_S$ on the basis of (${\rm Re} H^0_1$, ${\rm Re} H^0_2$, ${\rm Re} \xi^0$) as
\begin{equation}
M_S = \left ( \begin{array}{ccc}
    M_{S11} & M_{S12} & M_{S13}   \cr
    M_{S12} & M_{S22} & M_{S23}   \cr
    M_{S13} & M_{S23} & M_{S33}
        \end{array}
    \right ) \ ,
\end{equation}
its matrix elements at the one-loop level are calculated as follows:
\begin{eqnarray}
M_{S11} & = & m_Z^2 \cos^2 \beta  + M_{P11} \sin^2 \beta
- {3 \cos^2 \beta \over 16 \pi^2 v^2}
\bigg ({4 \over 3} m_W^2 - {5 \over 6} m_Z^2 \bigg )^2
f(m_{{\tilde t}_1}^2, m_{{\tilde t}_2}^2)  + M_{S11}^t \ ,  \cr
M_{S22} & = & m_Z^2 \sin^2 \beta + M_{P11} \cos^2 \beta
- {3 m_t^4 \over 4 \pi^2 v^2 \sin^2 \beta} \log \bigg ({m_t^2 \over \Lambda^2} \bigg )   \cr
& &\mbox{} - {3 \sin^2 \beta \over 16 \pi^2 v^2}
\bigg ({4 \over 3} m_W^2 - {5 \over 6} m_Z^2 \bigg )^2
f(m_{{\tilde t}_1}^2, m_{{\tilde t}_2}^2)  + M_{S22}^t    \ ,  \cr
M_{S33} & = & 4 B_T \mu_T + M_{P22}  + M_{S33}^t    \ , \cr
M_{S12} & = & {1 \over 2} (\lambda^2 v^2 - m_Z^2 - M_{P11}) \sin 2 \beta + M_{S12}^t   \ , \cr
M_{S13} & = & \lambda^2 v x \cos \beta
- {1 \over \sqrt{2}} \lambda A_{\lambda} v  \sin \beta
- \sqrt{2} \lambda \mu_D v  \cos \beta  - \sqrt{2} \lambda \mu_T v  \sin \beta   \cr
& &\mbox{} + {3 m_t^2 \lambda \cos \beta \over 32 \pi^2 v \sin^2 \beta}
(2 \sqrt{2} \mu_D - \sqrt{2} A_t - 2 \lambda x)
f(m_{{\tilde t}_1}^2, m_{{\tilde t}_2}^2)  + M_{S13}^t      \   , \cr
M_{S23} & = & \lambda^2 v x \sin \beta
- {1 \over \sqrt{2}} \lambda A_{\lambda} v  \cos \beta
- \sqrt{2} \lambda \mu_D v  \sin \beta - \sqrt{2} \lambda \mu_T v  \cos \beta \cr
& &\mbox{} - {3 \sqrt{2} A_t m_t^2 \lambda \over 32 \pi^2 v \tan \beta \sin \beta}
f(m_{{\tilde t}_1}^2, m_{{\tilde t}_2}^2)   + M_{S23}^t     \   ,
\end{eqnarray}
where $M_{Sij}^t$ ($i,j = 1,2,3$) come from the one-loop contributions.
Explicitly, they are given as
\begin{eqnarray}
M_{Sij}^t & = & {3 \over 32 \pi^2 v^2} W_i^C W_j^C
{g(m_{{\tilde t}_1}^2, m_{{\tilde t}_2}^2) \over (m_{{\tilde t}_2}^2 - m_{{\tilde t}_1}^2)^2}
+ {3 \over 32 \pi^2 v^2} A_i^C A_j^C
\log \left ( {m_{{\tilde t}_1}^2 m_{{\tilde t}_2}^2 \over \Lambda^4 } \right ) \cr
& &\mbox{} + {3 \over 32 \pi^2 v^2} (W_i^C A_j^C + A_i^C W_j^C)
{ \log ( m_{{\tilde t}_2}^2/ m_{{\tilde t}_1}^2)  \over (m_{{\tilde t}_2}^2 - m_{{\tilde t}_1}^2)}
\end{eqnarray}
where
\begin{eqnarray}
A_1^C & = & {m_Z^2 \over 2} \cos \beta \ , \cr
A_2^C & = & {2 m_t^2 \over \sin \beta} - {m_Z^2 \over 2} \sin \beta  \ , \cr
A_3^C & = & 0   \ ,  \cr
W_1^C & = & {m_t^2 \Delta_1^C \over \sin \beta } + \cos \beta \Delta_g  \ , \cr
W_2^C & = & {m_t^2 A_t \Delta_2^C \over \sin \beta } - \sin \beta \Delta_g   \ , \cr
W_3^C & = & {m_t^2 \lambda v \Delta_2^C \over \sqrt{2} \tan \beta}  \ ,
\end{eqnarray}
with
\begin{eqnarray}
\Delta_1^C & = & A_t (\sqrt{2} \lambda x - 2 \mu_D)
+ (2 \mu_D^2 + \lambda^2 x^2 - 2 \sqrt{2} \lambda x \mu_D) \cot \beta    \ , \cr
\Delta_2^C & = & 2 A_t + (\sqrt{2} \lambda x - 2 \mu_D) \cot \beta     \   ,  \cr
\Delta_g & = & \bigg ({4 \over 3} m_W^2 - {5 \over 6} m_Z^2 \bigg )
\bigg (m_Q^2 - m_U^2 + \bigg ({4 \over 3} m_W^2 - {5 \over 6} m_Z^2 \bigg) \cos 2 \beta \bigg)   \  ,
\end{eqnarray}
and
\begin{equation}
g(m_x^2,m_y^2) = {m_y^2 + m_x^2 \over m_x^2 - m_y^2} \log {m_y^2 \over m_x^2} + 2 \ .
\end{equation}
The analytic formulae for eigenvalues and eigenvectors of $M_S$ may be obtained
by using some mathematical techniques [16].

\subsection{CP-violating scenario}

Next, we consider the CP-violating scenario, where we assume that neither $\phi_1$ nor $\phi_2$ may
be zero in the vacuum expectation values of $H^0_2$ and $\xi^0$.
The five neutral Higgs bosons may not have definite CP parities as they are inevitably mixed.
In this scenario, we have the stop quark mixing term $X_t$ in the expression for the masses of the stop quarks as
\begin{eqnarray}
X_t & = & \left [ {m_Q^2 - m_U^2 \over 2} + \left ( {2 m_W^2 \over 3} - {5 m_Z^2 \over 12} \right ) \cos 2 \beta
   \right ]^2
   +  m_t^2 \bigg [A_t^2 - 2 A_t \mu_D \cos \phi_1 \cot \beta               \cr
& &\mbox{} + \sqrt{2} A_t \lambda x \cos \phi_1 \cos \phi_2 \cot \beta
   - \sqrt{2} \mu_D \lambda x \cos \phi_2 \cot^2 \beta \cr
& &\mbox{} + \mu_D^2 \cot^2 \beta + \lambda^2 x^2 \cot^2 \beta/2
   -\sqrt{2} \lambda x A_t \sin \phi_1 \sin \phi_2 \cot \beta \bigg ]  \  .
\end{eqnarray}

Note that the CP violating vacuum in this scenario is defined as the stationary point
with respect to the two complex phases $\phi_1$ and $\phi_2$.
In other words, the CP violating vacuum should satisfy two minimum conditions for $\phi_1$ and $\phi_2$.
These two minimum conditions may be used to eliminate two free parameters, which we take $B_D$ and $B_T$.
From the minimum equations for $\phi_1$ and $\phi_2$, respectively, we replace $B_D$ and $B_T$ by
\begin{eqnarray}
B_D & = & {\sqrt{2} \lambda \mu_T x \sin (\phi_1 - \phi_2) \over \mu_D \sin \phi_1}
+ {\lambda A_{\lambda} x \sin (\phi_1 + \phi_2) \over \sqrt{2} \mu_D \sin \phi_1} \cr
& &\mbox{} - {3 m_t^2 A_t \over 16 \pi^2 v^2 \sin^2 \beta}  f(m_{{\tilde t}_1}^2, \ m_{{\tilde t}_2}^2)
+ { 3 m_t^2 A_t \lambda x \sin (\phi_1 + \phi_2) \over 16 \sqrt{2} \pi^2 v^2 \sin^2 \beta \mu_D \sin \phi_1}
f(m_{{\tilde t}_1}^2, \ m_{{\tilde t}_2}^2)     \   ,    \cr
B_T & = &\mbox{} - {\lambda v^2 \sin 2 \beta \sin (\phi_1 - \phi_2) \over 2 \sqrt{2} x \sin (2 \phi_2)}
+ {\lambda v^2 \mu_D \over 4 \sqrt{2} \mu_T x \cos \phi_2} \cr
& &\mbox{} + {\lambda A_{\lambda} v^2 \sin 2 \beta \sin (\phi_1 + \phi_2) \over 4 \sqrt{2} \mu_T x \sin (2 \phi_2) }
- {3 m_t^2 \mu_D \lambda \cot^2 \beta \sin (\phi_2) \over 32 \sqrt{2} \pi^2 x \mu_T \sin (2 \phi_2)}
f(m_{{\tilde t}_1}^2, \ m_{{\tilde t}_2}^2) \cr
& &\mbox{} + { 3 m_t^2 A_t \lambda \cot \beta \sin (\phi_1 + \phi_2) \over 32 \sqrt{2} \pi^2 \mu_T x \sin (2 \phi_2)}
f(m_{{\tilde t}_1}^2, \ m_{{\tilde t}_2}^2)   \  .
\end{eqnarray}

The five neutral Higgs bosons, $h_i$ ($i = 1, 2, 3, 4, 5$), are constructed as linear combinations of
${\rm Re} H^0_1$, ${\rm Re} H^0_2$, ${\rm Re} \xi^0$, $\sin\beta {\rm Im}H^0_1 + \cos\beta {\rm Im} H^0_2$
and ${\rm Im} \xi^0$.
Their squared masses at the one-loop level, $m_{h_i}$ ($i = 1, 2, 3, 4, 5$),
are given as the eigenvalues of the symmetric $5\times 5$ mass matrix $M$ for them.
These neutral Higgs bosons are sorted such that $m_{h_i} < m_{h_j}$ for $i < j$.
Explicit calculations to obtain the eigenvalues and eigenvectors of $M$ are numerically
carried out by using CERN library program.

Let us write down $M$ for convenience as
\begin{equation}
    M = M^0 + M^t
\end{equation}
where $M^0$ represents the mass matrix for the neutral Higgs bosons at the tree level,
obtained from $V_0$, and $M^t$ is the one-loop contributions, obtained from $V_1$.
The explicit expressions for the matrix elements of $M^0$ and $M^t$ are obtained
on the basis of (${\rm Re} H^0_1$, ${\rm Re} H^0_2$, ${\rm Re} \xi^0$,
$\sin\beta {\rm Im}H^0_1 + \cos\beta {\rm Im} H^0_2$, ${\rm Im} \xi^0$) as follows:
For $M^0_{ij}$, we have
\begin{eqnarray}
M_{44}^0 & = & {\sqrt{2} \over \sin 2 \beta} \lambda A_{\lambda} x \cos (\phi_1 + \phi_2)
- {2 B_D \mu_D \cos \phi_1 \over \sin 2 \beta}
+ {2 \sqrt{2} \over \sin 2 \beta } \lambda \mu_T x \cos (\phi_1 - \phi_2)   \  ,  \cr
M_{55}^0 & = & {1 \over 2 \sqrt{2} x} \lambda A_{\lambda} v^2 \sin 2 \beta \cos (\phi_1 + \phi_2)
- 4 B_T \mu_T \cos (2 \phi_2) + {1 \over \sqrt{2} x } \lambda \mu_D v^2 \cos \phi_2    \cr
& &\mbox{} + {1 \over \sqrt{2} x} \lambda v^2 \mu_T \sin 2 \beta \cos (\phi_1 -\phi_2) \ , \cr
M_{11}^0 & = & m_Z^2 \cos^2 \beta + \sin^2 \beta M_{44}^0    \ ,  \cr
M_{22}^0 & = & m_Z^2 \sin^2 \beta + \cos^2 \beta M_{44}^0    \ ,  \cr
M_{33}^0 & = & 4 B_T \mu_T \cos (2 \phi_2) + M_{55}^0    \ , \cr
M_{12}^0 & = & {1 \over 2} (\lambda^2 v^2 - m_Z^2 - M_{44}^0) \sin 2 \beta     \ , \cr
M_{13}^0 & = & \lambda^2 v x \cos \beta
- {1 \over \sqrt{2}} \lambda A_{\lambda} v \cos (\phi_1 + \phi_2) \sin \beta
- \sqrt{2} \lambda \mu_D v \cos \phi_2 \cos \beta   \cr
& &\mbox{} - \sqrt{2} \lambda \mu_T v \cos (\phi_1 - \phi_2) \sin \beta \ , \cr
M_{14}^0 & = & 0 \  , \cr
M_{15}^0 & = & {1 \over \sqrt{2}} \lambda A_{\lambda} v \sin (\phi_1 + \phi_2) \sin \beta
+ \sqrt{2} \lambda \mu_D v \cos \beta \sin \phi_2  \cr
& &\mbox{} - \sqrt{2} \lambda \mu_T v \sin (\phi_1 - \phi_2) \sin \beta \ , \cr
M_{23}^0 & = & \lambda^2 v x \sin \beta
- {1 \over \sqrt{2}} \lambda A_{\lambda} v \cos (\phi_1 + \phi_2) \cos \beta
- \sqrt{2} \lambda \mu_D v \cos \phi_2 \sin \beta   \cr
& &\mbox{} - \sqrt{2} \lambda \mu_T v \cos (\phi_1 - \phi_2) \cos \beta \ , \cr
M_{24}^0 & = &  0  \   , \cr
M_{25}^0 & = & {1 \over \sqrt{2}} \lambda A_{\lambda} v \sin (\phi_1 + \phi_2) \cos \beta
+ \sqrt{2} \lambda \mu_D v \sin \beta \sin \phi_2  \cr
& &\mbox{} - \sqrt{2} \lambda \mu_T v \sin (\phi_1 - \phi_2) \cos \beta \ , \cr
M_{34}^0 & = & {1 \over \sqrt{2}} \lambda A_{\lambda} v \sin (\phi_1 + \phi_2)
+ \sqrt{2} \lambda \mu_T v \sin (\phi_1 - \phi_2) \ , \cr
M_{35}^0 & = &\mbox{} - 2 B_T \mu_T \sin (2 \phi_2) \ , \cr
M_{45}^0 & = & {1 \over \sqrt{2}} \lambda A_{\lambda} v \cos (\phi_1 + \phi_2)
- \sqrt{2} \lambda \mu_T v \cos (\phi_1 - \phi_2) \ .
\end{eqnarray}
and, for $M^t_{ij}$, we have
\begin{eqnarray}
M_{ij}^t & = & {3 \over 32 \pi^2 v^2} W_i^t W_j^t
{g(m_{{\tilde t}_1}^2, m_{{\tilde t}_2}^2) \over (m_{{\tilde t}_2}^2 - m_{{\tilde t}_1}^2)^2}
+ {3 \over 32 \pi^2 v^2} A_i^t A_j^t
\log \left ( {m_{{\tilde t}_1}^2 m_{{\tilde t}_2}^2 \over \Lambda^4 } \right ) \cr
& &\mbox{} + {3 \over 32 \pi^2 v^2} (W_i^t A_j^t + A_i^t W_j^t)
{ \log ( m_{{\tilde t}_2}^2/ m_{{\tilde t}_1}^2)  \over (m_{{\tilde t}_2}^2 - m_{{\tilde t}_1}^2)} + D_{ij}^t  \ ,
\end{eqnarray}
where
\begin{eqnarray}
A_1^t & = & {m_Z^2 \over 2} \cos \beta \ , \cr
A_2^t & = & {2 m_t^2 \over \sin \beta} - {m_Z^2 \over 2} \sin \beta  \ , \cr
A_3^t & = & 0   \ ,  \cr
A_4^t & = & 0   \ , \cr
A_5^t & = & 0   \ , \cr
W_1^t & = & {m_t^2 \Delta_1^t \over \sin \beta } + \cos \beta \Delta_g  \ , \cr
W_2^t & = & {m_t^2 A_t \Delta_2^t \over \sin \beta } - \sin \beta \Delta_g   \ , \cr
W_3^t & = & {m_t^2 \lambda v \Delta_3^t \over \sqrt{2} \tan \beta}  \ ,  \cr
W_4^t & = & {m_t^2 A_t \over \sin^2 \beta } \bigg ( 2 \mu_D \sin \phi_1
- \sqrt{2} \lambda x \sin (\phi_1 + \phi_2)  \bigg )   \ , \cr
W_5^t & = & {\sqrt{2} \lambda v m_t^2 \over \tan \beta }
\bigg [\mu_D \cot \beta \sin \phi_2 - A_t \sin (\phi_1 + \phi_2) \bigg ]   \ ,
\end{eqnarray}
with
\begin{eqnarray}
\Delta_1^t & = &\mbox{} - \sqrt{2} \lambda x A_t \sin \phi_1 \sin \phi_2 + A_t \cos \phi_1
(\sqrt{2} \lambda x \cos \phi_2 - 2 \mu_D)  \cr
& &\mbox{} + (2 \mu_D^2 + \lambda^2 x^2 - 2 \sqrt{2} \lambda x \mu_D \cos \phi_2) \cot \beta    \ , \cr
\Delta_2^t & = & 2 A_t + (\sqrt{2} \lambda x \cos \phi_2 - 2 \mu_D) \cot \beta \cos \phi_1
 - \sqrt{2} \lambda x \cot \beta \sin \phi_1 \sin \phi_2      \  ,  \cr
\Delta_3^t & = & 2 A_t \cos \phi_1 \cos \phi_2 + (\sqrt{2} \lambda x - 2 \mu_D \cos \phi_2) \cot \beta
- 2 A_t \sin \phi_1 \sin \phi_2     \ ,
\end{eqnarray}
and
\begin{eqnarray}
D_{44}^t & = &\mbox{}
- {3 m_t^2 A_t (2 \mu_D \cos \phi_1 - \sqrt{2} \lambda x \cos (\phi_1 + \phi_2) ) \over 32 \pi^2 v^2 \sin^3 \beta \cos \beta}
f(m_{{\tilde t}_1}^2, \ m_{{\tilde t}_2}^2) \ , \cr
D_{55}^t & = &\mbox{}
- {\sqrt{2} m_t^2 \lambda (\mu_D \cot \beta \cos \phi_2 - A_t \cos (\phi_1 + \phi_2) ) \over 32 \pi^2 x \tan \beta}
f(m_{{\tilde t}_1}^2, \ m_{{\tilde t}_2}^2) \ , \cr
D_{11}^t & = & \sin^2 \beta D_{44}^t - {3 \cos^2 \beta \over 16 \pi^2 v^2}
\bigg ( {4 m_W^2 \over 3} - {5 m_Z^2 \over 6} \bigg )^2
f(m_{{\tilde t}_1}^2, \ m_{{\tilde t}_2}^2) \ , \cr
D_{22}^t & = &  \cos^2 \beta D_{44}^t - {3 \sin^2 \beta \over 16 \pi^2 v^2}
\bigg ({4 m_W^2 \over 3} - {5 m_Z^2 \over 6} \bigg)^2
f(m_{{\tilde t}_1}^2, \ m_{{\tilde t}_2}^2)     \cr
& &\mbox{} - {3 m_t^4 \over 4 \pi^2 v^2 \sin^2 \beta} \log \bigg ({m_t^2 \over \Lambda^2} \bigg )  \ ,  \cr
D_{33}^t & = &  D_{55}^t  \ , \cr
D_{12}^t & = &\mbox{} - \cos \beta \sin \beta D_{44}^t + {3 \sin 2 \beta \over 32 \pi^2 v^2}
\bigg ({4 m_W^2 \over 3} - {5 m_Z^2 \over 6} \bigg )^2
f(m_{{\tilde t}_1}^2, \ m_{{\tilde t}_2}^2)  \ , \cr
D_{13}^t & = & {3 m_t^2 \lambda \cos \beta \over 32 \pi^2 v \sin^2 \beta}
\bigg (2 \sqrt{2} \mu_D \cos \phi_2 - \sqrt{2} A_t \tan \beta \cos (\phi_1 + \phi_2) \cr
& &\mbox{} - 2 \lambda x \bigg )
f(m_{{\tilde t}_1}^2, \ m_{{\tilde t}_2}^2)     \   ,    \cr
D_{14}^t & = & 0  \    ,    \cr
D_{15}^t & = &\mbox{} - {3 \sqrt{2} m_t^2 \lambda \over 32 \pi^2 v \sin \beta}
\bigg (2 \mu_D \cot \beta \sin \phi_2 - A_t \sin (\phi_1 + \phi_2) \bigg )
f(m_{{\tilde t}_1}^2, \ m_{{\tilde t}_2}^2)     \   ,    \cr
D_{23}^t & = &\mbox{} - {3 \sqrt{2} m_t^2 \lambda A_t \cos (\phi_1 + \phi_2) \over 32 \pi^2 v \tan \beta \sin \beta}
f(m_{{\tilde t}_1}^2, \ m_{{\tilde t}_2}^2)     \   ,    \cr
D_{24}^t & = & 0      \   ,     \cr
D_{25}^t & = & {3 \sqrt{2} m_t^2 \lambda A_t \sin (\phi_1 + \phi_2) \over 32 \pi^2 v \tan \beta \sin \beta}
f(m_{{\tilde t}_1}^2, \ m_{{\tilde t}_2}^2)     \   ,  \cr
D_{34}^t & = & {3 \sqrt{2} m_t^2 \lambda A_t \sin (\phi_1 + \phi_2) \over 32 \pi^2 v \sin^2 \beta}
f(m_{{\tilde t}_1}^2, \ m_{{\tilde t}_2}^2)     \   ,  \cr
D_{35}^t & = & 0   \   ,   \cr
D_{45}^t & = & {3 \sqrt{2} m_t^2 \lambda A_t \cos (\phi_1 + \phi_2) \over 32 \pi^2 v \sin^2 \beta}
f(m_{{\tilde t}_1}^2, \ m_{{\tilde t}_2}^2)     \   .
\end{eqnarray}

Here, it is worthwhile noting some points.
Note that $M_{14}^0 = M_{24}^0 = 0$. which implies that there is no mixing
between ${\rm Re} H^0_1$ and $\sin\beta {\rm Im}H^0_1 + \cos\beta {\rm Im} H^0_2$, nor between
${\rm Re} H^0_2$ and $\sin\beta {\rm Im}H^0_1 + \cos\beta {\rm Im} H^0_2$, at the tree level.
That is, there is no scalar-pseudoscalar mixings in the two Higgs doublets at the tree level.
They are mixed at the one-loop level due to the radiative corrections $M_{14}^t \neq 0$ and $M_{24}^t \neq 0$.

The spontaneous CP violation at the tree level is induced by $M_{15}$, $M_{25}$, $M_{34}$, and $M_{35}$,
among the two Higgs doublets and the Higgs triplet.
In particular, a self mixing in the Higgs triplet is represented by $M_{35}$.
If $\phi_1 = \phi_2 = 0$, all of these mixing terms would naturally disappear,
and consequently the $5 \times 5$ mass matrix for the neutral Higgs bosons would be decomposed
into a $3 \times 3$ and a $2 \times 2$ submatrices.

\section{Numerical analysis}

Now, we are interested in whether the TESSM at the one-loop level may have
a reasonable parameter space to allow spontaneous CP violation.
In order to find out the possibility, we first set up the reasonable ranges for relevant parameters.
We take the top quark mass as 175 GeV and the renormalization scale as 300 GeV.
We assume that the lighter stop quark is heavier than the top quark.

The ratio of $\tan\beta = v_2 /v_1$ is allowed to vary from 1 to 30,
since the radiative corrections from the bottom and sbottom quark loops may be neglected in this range.
For the vacuum expectation value of the Higgs triplet,
we set the range as $0.5 < x {\rm ~(GeV)~} < 2.5$, where the upper bound is determined
by the experimental constraint on the $\rho$ parameter, and the lower bound is chosen
in order to avoid unnecessary singularities in the mass matrix for the neutral Higgs bosons, $M$,
where some terms are proportional to $1/x$.
The two complex phases, $\phi_1$ and $\phi_2$, are allowed to vary within the range between 0 and $\pi$.

The soft SUSY breaking parameters appearing in the radiative corrections at the one-loop level are
allowed within the range of $100 < m_Q, m_U, A_t < 1000$ GeV.
In this model, there are two $\mu$ parameter.
The dimensionful parameter $\mu_D$ which comes from the mixing between the two Higgs doublets
and $\mu_T$ which comes from the self mixing of the Higgs triplet are allowed to vary respectively
in the ranges of  $150 < \mu_D {\rm ~(GeV)~} < 500$ and $0 < \mu_T {\rm ~(GeV)~} < 500$.
Note that the lower bound on $\mu_D$ is determined by the present experimental constraints
on the chargino systems [18].

The allowed ranges for other parameters should be determined with care.
The quartic coupling coefficient $\lambda$ is important because the lightest neutral Higgs boson mass depends
critically on it.
In Ref. [4], the upper bound on $\lambda$ is calculated as a function of $h_t$,
the Yukawa coupling of the top quark, by employing renormalization group equation.
Thus, the upper bound on $\lambda$ may be expressed, through $m_t = h_t v \sin \beta$,
in terms of $\tan \beta$ and the top quark mass,
We see that the upper bound on $\lambda$ increases monotonically up to about 0.9
as $\tan \beta$ increase from 1 to 30.
Thus, we set $0 < \lambda < 0.9$.

The trilinear mass parameter $A_{\lambda}$ also deserves careful attentions.
It appears in a number of supersymmetric models such as
the next-to minimal supersymmetric standard model [19],
the minimal nonminimal supersymmetric standard model [20],
the U(1)-extended supersymmetric standard model [21],
and the secluded $U(1)'$-extended MSSM [22].
In these models, the trilinear term with $A_{\lambda}$ plays an important role
to increase the strength of the first-order electroweak phase transition for baryogenesis
in order to describe the asymmetry of the matter and antimatter.

Meanwhile, for the explicit CP violation scenario in various nonminimal supersymmetric models [13,14,23],
the trilinear mass parameter $A_{\lambda}$ is found to generate the non-trivial CP phase.
Thus, $A_{\lambda}$ is very important in nonminimal supersymmetric models in order to achieve the explicit CP violation.
Referring to the results of those studies, we set the allowed range as $100 < A_{\lambda} {\rm ~(GeV)~}< 1000$.

We are now left with two parameters $B_D$ and $B_T$.
In the CP-conserving scenario, they are free parameters.
We set $- 500 < B_D {\rm ~(GeV)~}< 0$ and $- 500 < B_T {\rm ~(GeV)~}< 500$,
because the electroweak symmetry breaking is favored in these ranges, in the CP conserving scenario.
On the other hand, in the spontaneous CP-violating scenario, these parameters may
be eliminated by means of vacuum stability conditions, as mentioned before.
In this case, we need not establish ranges for them {\it a priori}, since their values are determined
in terms of other parameters.
Nevertheless, in the spontaneous CP-violating scenario, we would like to select the values of other parameters
such that $B_D$ and $B_T$ should be in the above ranges.
In other words, when we examine the parameter space, it is an internal constraint among the relevant parameters
that they should yield $- 500 < B_D {\rm ~(GeV)~}< 0$ and $- 500 < B_T {\rm ~(GeV)~}< 500$,
in the CP-violating scenario.

We first study the CP-conserving scenario.
In this scenario, we calculate the mass of the lightest scalar Higgs boson, $m_{S_1}$,
for given $\tan \beta$, both at the tree level and at the one-loop level,
where the values of other parameters are randomly selected in the parameter space defined as $0 < \lambda < 0.9$,
$100 < A_{\lambda} {\rm ~(GeV)~}< 1000$, $0.5 < x {\rm ~(GeV)~}< 2.5$,
$150 < \mu_D {\rm ~(GeV)~}< 500$, and $0 < \mu_T {\rm ~(GeV)~}< 500$,
$- 500 < B_D {\rm ~(GeV)~}< 0$, $- 500 < B_T {\rm ~(GeV)~}< 500$, and $m_Q, m_U, A_t$ between 100 GeV and 1000 GeV.
The calculation is repeated for randomly varying parameter values within the parameter space.
Then, for given $\tan\beta$, the largest value of $m_{S_1}$ is entitled as the upper bound on $m_{S_1}$.

The results of our  numerical calculations in the CP-conserving scenario are shown in Fig. 1,
where the solid curve is the upper bound on $m_{S_1}$ at the tree level and the dashed curve is
the corresponding one at the one-loop level, as a function of $\tan\beta$.
We find that our results are qualitatively consistent with other studies.
One can easily notice in Fig. 1 that the upper bound on $m_{S_1}$ at the one-loop level is as large as 140 GeV.
Even at the tree level, we find that the upper bound on $m_{S_1}$ may reach about 100 GeV,
This tree-level behavior is quite different from the MSSM, mainly because the quartic coupling possesses
the gauge couplings as well as $\lambda$ in this model, and, furthermore,
there is a additional quartic coupling, such as the top Yukawa coupling,
when the radiative corrections are included at the one-loop level.

Next, we study the CP-violating scenario.
The parameter space is defined as $0 < \phi_1, \phi_2 < \pi$, $0 < \lambda < 0.9$,
$100 < A_{\lambda} {\rm ~(GeV)~}< 1000$, $0.5 < x {\rm ~(GeV)~}< 2.5$,
$150 < \mu_D {\rm ~(GeV)~}< 500$, and $0 < \mu_T {\rm ~(GeV)~}< 500$,
and $m_Q, m_U, A_t$ between 100 GeV and 1000 GeV.
Note that $B_D$ and $B_T$ are dependent parameters.
In this parameter space, we select randomly $10^5$ points.
Each point represents a particular set of parameter values of ($\phi_1$, $\phi_2$, $\lambda$,
$A_{\lambda}$, $x$, $\mu_D$, $\mu_T$, $m_Q$, $m_U$, $A_t$).

For each point, we first calculate the values of $m_{h_i}$ ($i = 1,2,3,4,5$) at the tree level.
We find that $m_{h_1}$, the mass of the lightest neutral Higgs boson, in the TESSM
in the CP-violating scenario at the tree level are calculated to be between about 25 and 35 GeV.
The masses of other neutral Higgs bosons are calculated to be
$35 < m_{h_2} {\rm ~(GeV)~}< 57$,
$78< m_{h_3} {\rm ~(GeV)~} < 87$,
$850 <m_{h_4} {\rm ~(GeV)~}< 1560$,
and $1550 < m_{h_5} {\rm ~(GeV)~} < 2400$.

We also calculate $g_{ZZh_i}^2$ ($i = 1,2,3$) at the tree level.
These values are to be compared with experimental results, $(g^{\rm max}_{ZZH})^2$.
Here, $g^{\rm max}_{ZZH}$ is the model-independent upper bound on the coupling coefficient
between Higgs boson and a pair of $Z$ bosons, and it is given as a function of the mass of the Higgs boson
that couples to the pair of $Z$ bosons.
Recently, it has been measured by the LEP collaborations at the 95\% confidence level [15].
Note that we do not calculate $g_{ZZh_4}^2$ and $g_{ZZh_5}^2$.
Since the masses of $h_4$ and $h_5$ are calculated to be much larger than 120 GeV,
they are not constrained by the LEP results.

The solid curve in Fig. 2(a) shows $(g^{\rm max}_{ZZH})^2$, obtained from the LEP data,
as a function of the Higgs mass.
If the value of $g_{ZZh_i}^2$ is calculated to be larger than $(g^{\rm max}_{ZZH})^2$,
we should reject it, since it is beyond the experimental upper bound.
In Fig. 2(a), one can see not only the solid curve but also three crowds of points,
which are the results of our calculation.
These points are ($m_{h_1}$, $g_{ZZh_1}^2$) in the upper left corner of the figure (represented by stars),
($m_{h_2}$, $g_{ZZh_2}^2$) in the upper center (circles), and ($m_{h_3}$, $g_{ZZh_3}^2$)
in the upper right corner (crosses).
One may notice that the number of points in Fig.2(a) is far smaller than $10^5$.
This is because some of the $10^5$ sets of parameter values randomly selected
in the parameter space yield unphysical results, such as negative masses.
These unphysical results are rejected, and the points in Fig.2(a) are the accepted ones.

Now, it is easy to notice in Fig. 2(a) that most of these points are above the solid curve,
implying that $g_{ZZh_i}^2$ is calculated to be larger than the experimental upper bound,
$(g^{\rm max}_{ZZH})^2$.
In particular, all of ($m_{h_3}$, $g_{ZZh_3}^2$) are located above the solid curve.

The implication of our calculation is quite clear.
There is no parameter set, out of $10^5$ sets, that yields $g_{ZZh_3}^2$ smaller than $(g^{\rm max}_{ZZH})^2$.
This implies either that $h_3$ with a mass of about 80 GeV should have been discovered
via $ZZh_3$ coupling at LEP experiments or that no such $h_3$ with such ($m_{h_3}$, $g_{ZZh_3}^2$) is allowed,
and the latter is reasonably acceptable.
Therefore, all of the $10^5$ parameter sets do not satisfy the experimental constraints set by LEP and
thus should be rejected.
Thus, it is completely fair to conclude that the whole parameter space under consideration of the TESSM
at the tree level is excluded by LEP with respect to the spontaneous CP violation,
This tree-level behavior of the TESSM is comparable to the NMSSM which also has no spontaneous CP violation
in the tree-level Higgs sector [12].

However, we find that the situation is substantially improved at the one-loop level.
We repeat the numerical calculations to obtain the values of $m_{h_i}$ and $g_{ZZh_i}^2$ ($i = 1,2,3,4,5$)
at the one-loop level, for randomly chosen $10^5$ points in the parameter space,
which is identical to the tree-level one.
The mass of the lightest neutral Higgs boson in the TESSM at the one-loop level
in the spontaneous CP-violating scenario is calculated to be between about 12 and 101 GeV.
The masses of the heavier neutral Higgs bosons at the one-loop level are calculated to be
$114 < m_{h_2} {\rm ~(GeV)~}< 135$,
$224< m_{h_3} {\rm ~(GeV)~} < 1000$,
$230 <m_{h_4} {\rm ~(GeV)~}< 2030$,
and $380 < m_{h_5} {\rm ~(GeV)~} < 2030$.

We then calculate $g_{ZZh_1}^2$, at the one-loop level.
It is not necessary to calculate other $g_{ZZh_i}^2$ ($i = 2,3,4,5$), since only $h_1$ is lighter than 120 GeV.
Our results for the one-loop level are shown in Fig. 2(b).
Here, a swarm of points are distributed over a large area of the ($m_{h_1}$, $g_{ZZh_1}^2$)-plane,
and well below the solid curve, which is $(g^{\rm max}_{ZZH})^2$ of the LEP experiments.
These points are all acceptable, as they satisfy the experimental constraints.
Therefore, we conclude that the parameter space of the TESSM at the one-loop level under consideration is
allowed by the LEP constraints allow for the spontaneous CP violation to take place.

\section{Conclusions}

We study the TESSM, where a chiral Higgs triplet with zero hypercharge is additionally introduced to the MSSM,
in order to examine the possibility of spontaneous CP violation in its Higgs sector.
This model possesses three charged Higgs bosons and five neutral Higgs bosons.
If the CP symmetry is conserved, the five neutral Higgs bosons have definite CP parities,
divided into three scalar and two pseudoscalar neutral Higgs bosons.
In this case, the upper bound on the mass of the lightest scalar Higgs boson is
about 103 GeV and 143 GeV at the tree level and at the one-loop level, respectively.

For the spontaneous CP violation to occur, we allow complex phases in
the vacuum expectation values of the Higgs doublets as well as the Higgs triplet.
Among them, two independent complex phases are introduced.
These complex phases induce the scalar-pseudoscalar mixings.
At the tree level, the scalar-pseudoscalar mixings take place between the Higgs doublets
and the Higgs triplet, but not between the two Higgs doublets.
However, at the one-loop level, the scalar-pseudoscalar mixings take place among them all.

We establish a reasonable parameter space in the TESSM, and,
for $10^5$ sets of relevant parameter values within the parameter space,
we calculate the masses of the five neutral Higgs bosons and their coupling coefficients $g_{ZZh_i}$
to a pair of $Z$ bosons, in the CP-violating scenario, at the tree level as well as at the one-loop level.
We find that $g_{ZZh_i}$ ($i = 1,2$) are calculated to exceed
the model-independent experimental upper bound set by LEP for nearly most of the parameter value sets,
and all of $g_{ZZh_3}$ are calculated to be larger than the experimental upper bound at the tree level.
Therefore, the parameter space of the TESSM for the spontaneous CP violation at the tree level is excluded
by the experimental constraint.
Practically, the spontaneous CP violation is impossible for the tree-level potential of the TESSM.

At the one-loop level, we find that $g_{ZZh_1}$ are calculated to stay within the experimental constraint LEP,
for the parameter space in consideration.
This implies that the spontaneous CP violation is possible in the TESSM at the one-loop level.
Meanwhile, the mass of the lightest neutral Higgs boson may be as small as 12 GeV in this case.
However, this does not contradict the negative result of Higgs search at LEP, since the Higgs couplings
to a $Z$ boson pair might also very small.

In conclusion, we confirm the possibility of spontaneous CP violation in the TESSM at the one-loop level.

\vskip 0.3 in
\noindent
{\large {\bf Acknowledgments}}
\vskip 0.2 in

S. W. Ham thanks his late teacher, Bjong Ro Kim, for learning supersymmetry.
He thanks Prof. D Son for the hospitality at KNU where a part of this work has been performed.
He also thanks Prof. P Ko for the hospitality at KIAS where a part of this work has been performed.
He is supported by the Korea Research Foundation Grant funded by the Korean Government
(MOEHRD, Basic Research Promotion Fund) (KRF-2007-341-C00010).
He was supported by grant No. KSC-2008-S01-0011 from Korea Institute of Science and Technology Information.
This work is supported by Konkuk University in 2007.

\vfil\eject


\vfil\eject
\setcounter{figure}{0}
\def\figurename{}{}%
\renewcommand\thefigure{FIG. 1}
\begin{figure}[t]
\begin{center}
\includegraphics[scale=0.6]{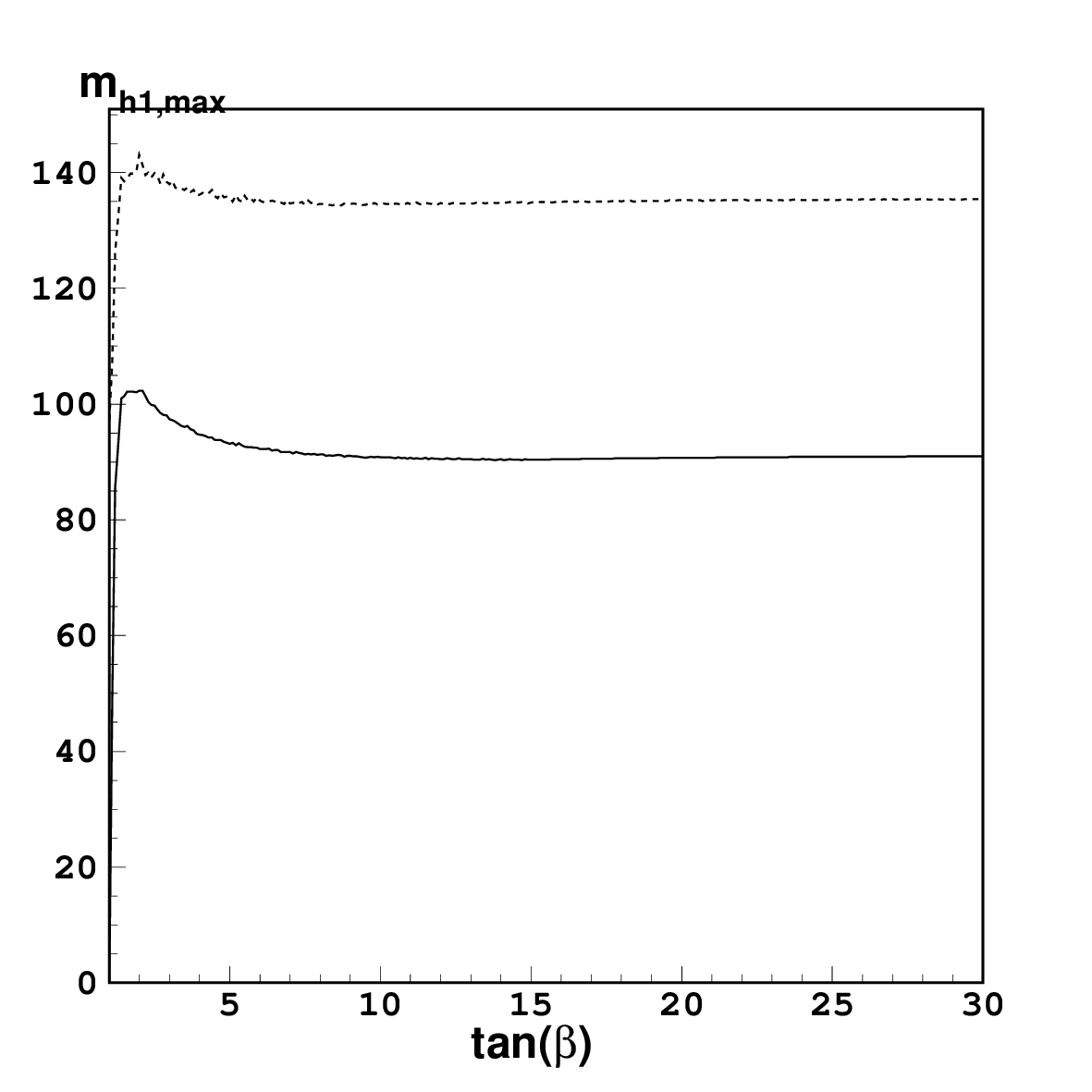}
\caption[plot]{
The upper bound on $m_{S_1}$ at the tree level (solid curve) and
at the one-loop level (dashed curve) as a function of $\tan \beta$, in CP-conserving scenario,
where the values of other parameters are randomly selected
in the parameter space defined as $0 < \lambda < 0.9$,
$100 < A_{\lambda} {\rm ~(GeV)~}< 1000$, $0.5 < x {\rm ~(GeV)~}< 2.5$,
$150 < \mu_D {\rm ~(GeV)~}< 500$, and $0 < \mu_T {\rm ~(GeV)~}< 500$,
$- 500 < B_D {\rm ~(GeV)~}< 0$, $- 500 < B_T {\rm ~(GeV)~}< 500$, and
$m_Q, m_U, A_t$ between 100 GeV and 1000 GeV.
Repeating the calculations for $10^5$ random points in the parameter space,
for given $\tan\beta$, the largest value of $m_{S_1}$  is defined as the upper bound on $m_{S_1}$.}
\end{center}
\end{figure}

\setcounter{figure}{0}
\def\figurename{}{}%
\renewcommand\thefigure{FIG. 2(a)}
\begin{figure}[t]
\begin{center}
\includegraphics[scale=0.6]{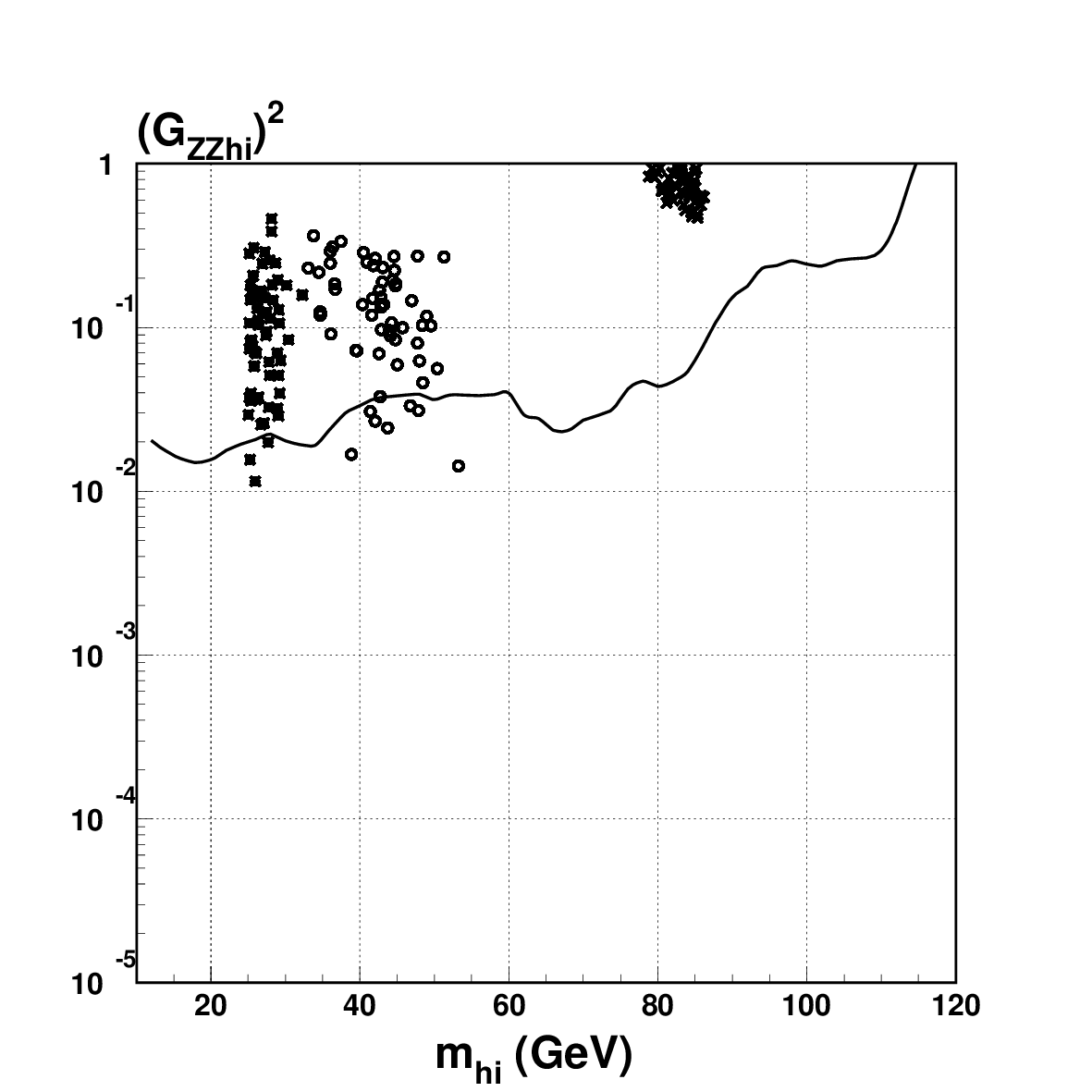}
\caption[plot]{Three crowds of points are ($m_{h_1}$, $g_{ZZh_1}^2$)
in the upper left corner of the figure (represented by stars),
($m_{h_2}$, $g_{ZZh_2}^2$) in the upper center (circles),
and ($m_{h_3}$, $g_{ZZh_3}^2$) in the upper right corner (crosses).
The parameter space is the same as in Fig.1.
The solid curve is the model-independent upper bound on $g_{ZZH}^2$,
the square of the coupling of a given Higgs boson to a pair of $Z$ bosons,
obtained from the LEP experiments, plotted as a function of the mass of the given Higgs boson.
Please notice that all points of ($m_{h_3}$, $g_{ZZh_3}^2$) are above the solid curve.}
\end{center}
\end{figure}

\setcounter{figure}{0}
\def\figurename{}{}%
\renewcommand\thefigure{FIG. 2(b)}
\begin{figure}[t]
\begin{center}
\includegraphics[scale=0.6]{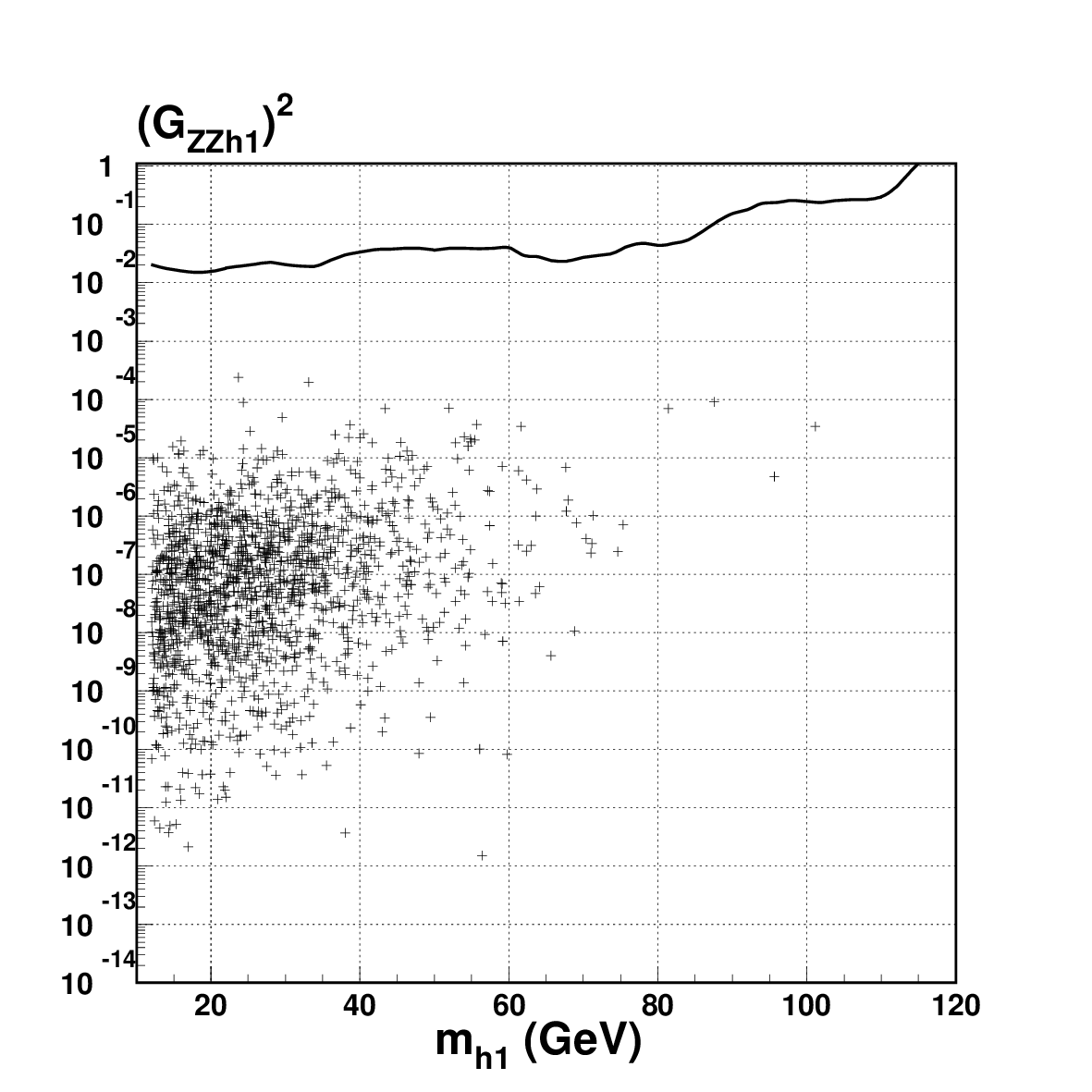}
\caption[plot]{A swarm of points are distributed in the ($m_{h_1}$, $g_{ZZh_1}^2$)-plane.
The parameter space is the same as in Fig.1, and the solid curve is the same as in Fig.2(a).
Please notice that all points of ($m_{h_1}$, $g_{ZZh_1}^2$) are below the solid curve. }
\end{center}
\end{figure}

\end{document}